\documentclass{revtex4}
\begin{document}
\hspace*{4.5 in}CUQM-125
\vspace{1 in}

\markboth{B Champion, R Hall, N Saad}{Asymptotic Iteration Method for singular potentials}
\title{Asymptotic Iteration Method for singular potentials}
\author{Brodie Champion}
\address{Department of Mathematics and Statistics,
University of Prince Edward Island,
550 University Avenue, Charlottetown,
Prince Edward Island, Canada C1A 4P3\\
bchampion@upei.ca}
\author{Richard L Hall}
\address{Department of Mathematics and Statistics, Concordia University,
1455 de Maisonneuve Boulevard West, Montr\'eal,
Qu\'ebec, Canada H3G 1M8\\
rhall@mathstat.concordia.ca}
\author{Nasser Saad}
\address{Department of Mathematics and Statistics,
University of Prince Edward Island,
550 University Avenue, Charlottetown,
Prince Edward Island, Canada C1A 4P3\\
nsaad@upei.ca}
\begin{abstract}
The asymptotic iteration method (AIM) is applied to obtain highly accurate eigenvalues of the radial Schr\"odinger equation with the singular potential $V(r)=r^2+\lambda/r^\alpha~~(\alpha,\lambda>0)$ in arbitrary dimensions.  Certain fundamental conditions for the application of AIM, such as a suitable asymptotic form for the wave function,  and the termination condition for the iteration process, are discussed. Several suggestions are introduced to improve the rate of convergence and to stabilize the computation. AIM offers a simple, accurate, and efficient method for the treatment of singular potentials such as $V(r)$ valid for all ranges of coupling $\lambda.$ 
\end{abstract}
\maketitle
\section{Introduction}	
Attractive potentials with a strong repulsive core are common in atomic, nuclear and molecular physics \cite{cas}$-$\cite{mz}. A typical class of such potentials \cite{har}$-$\cite{chs} are the spiked harmonic oscillators $V(r)=r^2+\frac{\lambda}{r^\alpha}$. The potential $V(r)$ is called `spiked' because of its pronounced peak near the origin. For $\alpha>2$, the potential is of relevance to quantum field theory, describing so called supersingular interactions for which matrix elements of the perturbation in the harmonic-oscillator basis diverge: thus every term in the perturbation series is infinite, and the perturbation expansion does not exist \cite{dk}. There are several other reasons for interest in this class of potentials. First, it represents the simplest example of certain class of realistic interaction models in atomic, nuclear and molecular physics. Second, in the one-dimensional case, the perturbed-oscillator operator $H=p^2+V(r)=-d^2/r^2+r^2+\lambda/r^\alpha=H_0+\lambda/r^\alpha$, where $p=-i\partial/\partial r$, may not converge to the original one $H_0$ as $\lambda\rightarrow 0$ (The Klauder phenomenon). Third, the perturbation series is ordered in fractional power \cite{har,shk} in $\lambda$. Since the early study of Harrell on singular perturbation theory for the ground-state energy of the Hamiltonian $H$, an enormous amount of work has been done to investigate the spectral problems of this operator. Most of the work, however, is either devoted to the study of the ground-state energy for $\lambda$ near zero, which represents the most challenging problem, or for particular values of the potential parameters. Owing to the difficulties inherited by the potential structure near the origin, much attention must be paid to the selected method for tackling these problems.  
The purpose of this letter is twofold. First to develop a simple and easily adopted technique, based on the asymptotic iteration method \cite{chs}, to compute the eigenvalues of the radial Schr\"odinger equation 
\begin{equation}\label{eq1}
-{d^2\psi(r)\over dr^2}+\bigg(r^2+{\gamma(\gamma+1)\over r^2}+{\lambda\over r^\alpha}\bigg)\psi(r)=E\psi(r), r\in [0,\infty),
\end{equation}
where $\alpha>0$, and the eigenfunctions $\{\psi(r)\}$ satisfy the Dirichlet boundary condition $\psi(0)=0$. This is
valid for the one-dimensional case as well as for higher dimensions $N > 1$ through $\gamma=l+{1\over 2}(N-3)$, regardless of the values of the potential parameters.
Second, to point out the importance of the correct form of asymptotic wave function to stabilize the iteration technique, and also to provide some suggestions to improve the rate of convergence of AIM when it is used to tackle Schr\"odinger eigenvalue problems with a wide variety of other singular potentials of physical and chemical interest. In order to achieve these goals, we develop first a wave function with the right exponential tail and which satisfies the Dirichlet boundary condition at the origin. This optimizes and stabalizes the use of AIM for computing the eigenvalues.
The asymptotic iteration method (AIM) was original introduced \cite{chs} to investigate the solutions of differential equations of the form 
\begin{equation}\label{eq2}
y''=\lambda_0(r) y'+s_0(r) y, \quad\quad\quad ({~}'={d\over dr})
\end{equation}
where $\lambda_0(r)$ and $s_0(r)$ are $C^{\infty}$-differentiable functions. Using AIM, the differential equation (\ref{eq2}) has a general solution \cite{chs}:
\begin{equation}\label{eq3}
y(r)= \exp\left(-\int\limits^{r}\rho(t) dt\right)
\left[C_2 +C_1\int\limits^{r}\exp\left(\int\limits^{t}(\lambda_0(\tau) + 2\rho(\tau)) d\tau \right)dt\right]
\end{equation}
where, for sufficiently large $n>0$, we obtain the $\rho(r)$ values from
\begin{equation}\label{eq4}
{s_{n}(r)\over \lambda_{n}(r)}={s_{n-1}(r)\over \lambda_{n-1}(r)} \equiv \rho(r)
\end{equation}
for
\begin{eqnarray}\label{eq5}
\lambda_{n}&=& \lambda_{n-1}^\prime+s_{n-1}+\lambda_0\lambda_{n-1}\nonumber\\
 s_{n}&=&s_{n-1}^\prime+s_0\lambda_{n-1},\quad n=1,2,3,\dots
\end{eqnarray}
It should be noted that one can start the iteration from $n=0$ with the initial condition $\lambda_{-1}=1$ and $s_{-1}=0$. Since Ref.\cite{chs} the method has been adopted to investigate the spectrum of many problems in relativistic and non-relativistic quantum mechanics \cite{chs}${-}$\cite{sh}. 
In the process of applying AIM, especially in its application to eigenvalue problems of Schr\"odinger-type, such as (\ref{eq1}), one usually faces the following two problems.
\subsection{Asymptotic solution Problem:} 
The first problem we are confronted with in applying AIM is the conversion of the eigenvalue problem (the absence of first derivative) to standard form suitable to utilize AIM (\ref{eq2}). A general strategy to overcome this problem is to use an asymptotic solution $\psi_{a}(r)$ which satisfies the boundary conditions of the given eigenvalue equation. By substitution of the assumed exact solution with the form $\psi(r)=\psi_{a}(r)f(r)$ into the eigenvalue problem, once the $\lambda_0$ and $s_0$ have been determined, the sequences $\lambda_n$ and $s_n$ can be computed using (\ref{eq5}). The energy eigenvalues are then obtained from the roots of the termination condition (\ref{eq4}), which can be written in more convenient form as  
\begin{equation}\label{eq6}
\delta_n(r;E)=\lambda_{n}(r;E) s_{n-1}(r;E)-\lambda_{n-1}(r;E) s_{n} (r;E)=0, \quad\quad n=1,2,\dots.
\end{equation}
For Schr\"odinger's equation (\ref{eq1}), the asymptotic solution is dominated by the harmonic oscillator wave function, since, for larger $r$, the dominant term of the potential is the harmonic oscillator term $r^2$. The problem with such asymptotic solution, however, is that the behavior of the wave function near the origin has not been considered. 
\subsection{Termination Condition Problem:} 
The second problem results when the eigenvalue problem (now in the standard form  for AIM application) fails to be exactly solvable. Indeed, if the eigenvalue problem has exact analytic solutions, the termination condition (\ref{eq6}) leads to an expression that depends only on the eigenvalues $E$, that is to say, independent of $r$. In such cases, the energy eigenvalues are simply the roots of $\delta_n(E)=0, n=1,2,\dots$. For example, if $\alpha=2$, Eq.(\ref{eq1}) is exactly solvable, and the termination condition (\ref{eq6}) yields
\begin{equation}\label{eq7}
\delta_n(E)=\prod_{i=0}^n(4i+3+2\gamma-E),\quad n=0,1,2,3,\dots.
\end{equation}
The condition $\delta_n(E)=0$ leads to the exact solutions $E_n=4n+3+2\gamma$, $n=0,1,2,\dots$ as expected \cite{chs}. If the eigenvalue problem is not analytically solvable with the analytic form chosen, as for $0<\alpha\neq 2$, then the termination condition (\ref{eq6}) produces for each iteration an expression that depends on both $r$ and $E$. In such a case, one faces the problem of finding the best possible starting value $r=r_0$ that stabilizes the process so that it doesn't oscillate but converges. In principle, the computation of the roots of $\delta_n(r_0;E)=0$ should be independent of the choice of $r_0$, nevertheless, the right choice of $r$, as we shall show in the present work, usually accelerates the rate of convergence to accurate eigenvalues $E$ within a reasonable number of iterations. Generally, a suitable $r_0$ value is determined either as the location of the maximum of value of the asymptotic wave function, or as the position of the minimum value of the potential under consideration. A more general and systematic way to choose a suitable value for $r_0$ is still open question for further research. 
\vskip0.1true in
For the Schr\"odinger equation (\ref{eq1}), we face these two problems. In the next section we develope an asymptotic wave function that satisfies the boundary conditions at zero and infinity.  This asymptotic form is then used in section 3 to initialized the asymptotic iteration method. In section 4, using a suitable value of $r_0$,  we exhibit and discuss the numerical results of AIM for a wide range of $\lambda$ and $\alpha>0$. Finally, in section 5, we comment on these results.
\section{Asymptotic wavefunction for singular potentials}
For small $r$, one can neglect in (\ref{eq1}) the energy $E$ and the harmonic oscillator term as compared with the perturbative term $\lambda/r^\alpha$. Hence near the origin, (\ref{eq1}) can be written as
\begin{equation}\label{eq21}
-\frac{d^2\psi}{dr^2}+\left(\frac{\gamma(\gamma+1)}{r^2}+\frac{\lambda}{r^\alpha}\right)\psi=0.
\end{equation}
Using the transformation
$$\psi(r)=\sqrt{r}\phi(t),\quad t=\beta r^\sigma$$
a straightforward calculation shows that Eq.(\ref{eq21}) can be written as 
$$
\frac{1}{4r^{3/2}}\phi(t)-{\sigma^2\beta\over  r^{-\sigma+3/2}}\frac{d\phi}{dt}-{\sigma^2\beta^2\over  r^{-2\sigma+3/2}}\frac{d^2\phi}{dt^2}+\frac{\gamma(\gamma+1)}{r^{3/2}}\phi(t)+{\lambda \over r^{\alpha-1/2}}\phi(t)=0
$$
or, in more compact form, as
$$
\frac{d^2\phi}{dt^2}+\frac{1}{t}\frac{d\phi}{dt}-\left[\frac{(2\gamma+1)^2}{4\sigma^2t^2}+\frac{\lambda}{\sigma^2\beta^2}
r^{-\alpha+2-2\sigma}\right]\phi=0.
$$
Therefore, for $2\sigma=2-\alpha$ and $\beta=\sqrt{\lambda}/\sigma$, we obtain the modified Bessel's differential equation
\begin{equation}\label{eq23}
\frac{d^2\phi}{dt^2}+\frac{1}{t}\frac{d\phi}{dt}-\left[\frac{\nu^2}{t^2}+1\right]\phi=0,
\end{equation}
where $\nu=\frac{2\gamma+1}{2-\alpha}$. The general solution of this differential equation is \cite{as}
$$
\phi(t)=c_1I_\nu(t)+c_2K_\nu(t),
$$
where $c_1$ and $c_2$ are constants and $I_\nu$ and $K_\nu$ are the modified Bessel functions of the first and second kind respectively \cite{as}. Consequently, an asymptotic solution of (\ref{eq1}) for zero energy is given by
\begin{equation}\label{eq24}
\psi(r)=\sqrt{r}[c_1I_\nu(\beta r^\sigma)+c_2K_\nu(\beta r^\sigma)].
\end{equation}
We may now consider two cases, depending on the value of $\alpha$.
\vskip0.1true in
\noindent {\bf Case I ($\alpha>2$):} In this case $\sigma < 0$, the boundary condition $\psi(0)=0$ forces $c_1=0$, hence
\begin{equation}\label{eq25}
\psi_{a}(r)\equiv c_2\cdot\sqrt{r}K_\frac{2\gamma+1}{\alpha-2}(\frac{2\sqrt{\lambda}}{\alpha-2} r^{1-\frac{\alpha}{2}}).
\end{equation}
From the asymptotic approximation  \cite{as} of $K_\nu(z)$, we know for large argument $z$ that $K_\nu(z)\equiv {e^{-z}}/{\sqrt{\frac{2z}{\pi}}}$. Therefore, since ${1-\frac{\alpha}{2}}<0$, we have for small $r$, that 
\begin{equation}\label{eq26}
\psi_{a}(r)\equiv c_1\cdot{\frac{1}{2}}\sqrt{\frac{\pi(\alpha-2)}{\sqrt{\lambda}}}r^{\alpha/4}e^{-\frac{2\sqrt{\lambda}}{\alpha-2}r^{1-\frac{\alpha}{2}}}.
\end{equation}
Consequently, for $\alpha=2m+2$, $m>0$, 
\begin{equation}\label{eq27}
\psi_{a}(r)\equiv  r^{(m+1)/2}e^{-\frac{\sqrt{\lambda}}{m r^{m}}}, \quad m> 0
\end{equation}
up to a constant. In particular, if $\alpha=4,$ (i.e. $m =1$), one recovers the familiar limiting form of the solution when the repulsive potential is proportional to $r^{-4}$:
$$
\psi_{a}(r)\equiv re^{-\frac{\sqrt{\lambda}}{ r}}\quad\hbox{as}\quad r\rightarrow 0.
$$
\noindent{\bf Case II ($0<\alpha<2$)}: In this case $\sigma>0$, the boundary condition $\psi(0)=0$ forces $c_2=0$ in (\ref{eq24}), thus
\begin{equation}\label{eq28}
\psi_{a}(r)\equiv c_2\cdot\sqrt{r}I_\frac{2\gamma+1}{2-\alpha}(\frac{2\sqrt{\lambda}}{2-\alpha} r^{1-\frac{\alpha}{2}}).
\end{equation}
For small argument $z$, the asymptotic approximation  \cite{as} of  
$I_\nu(z)\equiv \left(\frac{z}{2}\right)^\nu/\Gamma(\nu+1)$ yields
\begin{equation}\label{eq29}
\psi_{a}(r)\equiv r^{\gamma+1}
\end{equation}
up to a constant. 
\section{Applications}
Depending on the degree of singularity of the potential at the origin, that is characterized by 
the positive parameter $\alpha$, we have the following two cases:
\subsection{The case $\alpha=2m+2,~~ m>0$:} 
In this case the Hamiltonian operator 
\begin{equation}\label{eq31}
H=-{d^2\over dr^2} +r^2+{\gamma(\gamma+1)\over r^2} +{\lambda\over r^\alpha},\quad\quad \alpha>2
\end{equation}
leads to a non-Fuchsian singularity  \cite{ge} at $r=0$ of the Schr\"odinger equation $H\psi=E\psi$, because the potential term  $\lambda/r^\alpha$ has a pole of order $>2$. The asymptotic wave function developed in the previous section suggests that the exact solution of (\ref{eq1}), in the case $\alpha>2$, takes the form
\begin{equation}\label{eq32}
\psi(r)=r^{m+1\over 2} e^{-{r^2\over 2}-{\sqrt{\lambda}\over m r^m}}f(r).
\end{equation}
The first exponential term in (\ref{eq32}) takes into account the fact that for large $r$, the term $r^2$ in (\ref{eq31}) dominates over all other terms of the potential, including, of course, $\gamma(\gamma+1)/r^2$. For this wave function, Schr\"odinger's equation (\ref{eq1}) now reads
\begin{equation}\label{eq33}
f''(r)=\bigg(2r-{2\sqrt{\lambda}\over r^{m+1}}-{1+m\over r}\bigg)f'(r)+\bigg(2+m+
{2\sqrt{\lambda}\over r^{m}}+{(2\gamma+1-m)(2\gamma+1+m)\over 4r^2}-E\bigg)f(r),
\end{equation}
which is now amenable to AIM applications. Here, the primes of $f(r)$ in (\ref{eq33}) denote the derivatives with respect to $r$. 
\subsection{The case $\alpha=2m+2,~~ -1<m<0$:} 
In this case, the exact solution of (\ref{eq1}) assumes, for $\alpha<2$, the form
\begin{equation}\label{eq34}
\psi(r)=r^{\gamma+1}e^{-{r^2\over 2}}f(r),
\end{equation}
where again the exponential term in (\ref{eq34}) takes into account that for large $r$, the term $r^2$ dominates over all other terms of the potential. In this case, Schr\"odinger's equation (\ref{eq1}) reads
\begin{equation}\label{eq35}
f''(r)=2\bigg(r-{\gamma+1\over r}\bigg)f'(r)+\bigg(2\gamma+3+{\lambda\over r^{\alpha}}-E\bigg)f(r),
\end{equation}
which is suitable for an AIM application. 
\section{Iterative solutions}
For a given $\alpha$, using (\ref{eq33}) or (\ref{eq35}), we can explicitly write $\lambda_0(r)$ and $s_0(r)$ as:
\begin{itemize}
\item $\alpha>2$:
\begin{equation}\label{eq41}\left\{ \begin{array}{ll}
 \lambda_0(r)=\bigg(2r-{2\sqrt{\lambda}\over r^{m+1}}-{1+m\over r}\bigg), \\
  s_0(r)=\bigg(2+m+
{2\sqrt{\lambda}\over r^{m}}+{(2\gamma+1-m)(2\gamma+1+m)\over 4r^2}-E\bigg)
       \end{array} \right.
\end{equation}
where $m={1\over 2}(\alpha-2)$.
\item $\alpha<2$:
\begin{equation}\label{eq42}
\left\{ \begin{array}{ll}
\lambda_0(r)=2\bigg(r-{\gamma+1\over r}\bigg), \\
 s_0(r)=\bigg(2\gamma+3+{\lambda\over r^{\alpha}}-E\bigg)
       \end{array} \right.
\end{equation}
\end{itemize}
and, by means of the iteration formulas (\ref{eq5}), we calculate $\lambda_n(r)$ and $s_n(r)$, $n=1,2,\dots$. The eigenvalues are then computed by means of the termination condition (\ref{eq6}), namely $\delta_n(r;E)=0$. 
\begin{table}[h]
\caption{The effect of using different values of $r=r_0$ on computing the eigenvalues using AIM for the Schr\"odinger equation $-\frac{d^2\psi}{dr^2}+\left(r^2+\frac{0.1}{r^4}\right)\psi=E\psi$. An accurate value (to seven figures) is $E_{\rm exact}=3.575~552$. Here, $N$ refers to the number of iterations needed to achieve such accuracy.}
{\begin{tabular}{|r||c|c|c|c|c|}
\hline
  &  \multicolumn{5}{c|}{$r_0$}  \\
\hline
$N$ &  1 & 2 & 3 & 4 & 5 \\
\hline
\hline
15 & 3.478854 & 3.574644 & 3.570009 & 3.561477 & 3.555018 \\ 
20 & Fails & 3.575335 & 3.573786 & 3.569455 & 3.562441 \\ 
40 & ``" & 3.575551 & 3.575505 & 3.575298 & 3.574726 \\ 
60 & ``"      &    Fails      & 3.575549 & 3.575529 & 3.575458 \\ 
85 & ``"      & ``"& 3.575552 & 3.575550 & 3.575542 \\ 
90 & ``"      & ``" & Done      & 3.575551 & 3.575545 \\ 
115 & ``"      & ``"      & ``"     & 3.575552 & 3.575551 \\ 
120 & ``"      & ``"       & ``"    & Done     & 3.575551 \\ 
\hline
\end{tabular}}
\end{table}
With several symbolic mathematical programs available ({\it Maple, Mathematica,} etc), the computation of the eigenvalues by means of the iteration method, provided it is set up correctly, is a straightforward calculation, even for the higher iteration steps. Most of our computations in the present work were done using {\it Maple}  version 9 running on an IBM architecture personal computer (Dell Dimension 4400). As we mentioned above, the computation of the eigenvalues by means of (\ref{eq6}) should be independent of the choice of $r$.
However, in some applications, for certain values of $r$, we may encounter oscillations of the computed roots that seem to diverge in behavior. This is presumably due to rounding and computational errors in the algorithms used. In Table 1, we show the effects of choosing different starting values of $r\equiv r_0$ on the number of iterations. It is clear that $r_0\geq 3$ is sufficient as starting value of $r$. However, suitable choices for $r_0$ can significantly reduce the number of iterations needed to achieve the required accuracy.
In many cases, we have removed the oscillating behavior by increasing the number of significant digits that {\it Maple} uses in numerical computations. In Table 2, we illustrate the effect of using different numbers of significant digits on the iteration convergence using a {\it Maple} environment. As indicated by the results in the table, a higher-precision environment can remove the oscillation behavior, as well as stabilize the numerical computation of the root problem by means of the iteration process. In order to accelerate the computation we have written our code for root-finding algorithm instead of using the default procedure {\ttfamily Solve} of {\it Maple.}  
\begin{table}[h]
\caption{The effect of using different number of digits (in {\it Maple 9}) when computing the eigenvalues using AIM for the Hamiltonian $-\frac{d^2}{dr^2}+r^2+\frac{6}{r^2}+\frac{0.1}{r^3}$. Here we set $r=3$ in (\ref{eq6}). An accurate  value is $E_{\rm exact}=7.029816$. Here $N$ refers to the number of iterations.}
{\begin{tabular}{|r||c|c|c|c|}
\hline
&  \multicolumn{4}{c|}{Number of Digits} \\
\hline
$N$ &    10    &    14    &     18 & 22    \\
\hline
\hline
70 & 7.0301659550 & 7.0298160740 & 7.0298162024 & 7.0298162024 \\ 
75 & 7.0298754200 & 7.0298162383 & Done          & Done \\ 
80 & 7.0298152350 & 7.0298162511 &  ``"         &  ``"         \\ 
85 & 7.0298160500 & 7.0298162636 &  ``"         &  ``"         \\ 
90 & 7.0298162400 & Done         &  ``"         &  ``"         \\ 
\hline
\end{tabular}}
\end{table}
\noindent For values of $\alpha=1$, AIM gives excellent results, even for extremely small value of the coupling parameter $\lambda$. In table 3, we report the AIM results for a considerable  range of $\lambda$ values. These computations have been made with $r_0=3$. For fractional $\alpha$, such as $1/2, 3/2, 1.9$ etc., the method is still stabile and works well; however the number of iterations is much larger than that needed for integer $\alpha.$ The eigenvalues reported in Table 3 are in excellent agreement with the exact eigenvalue computed by means of a numerical integration of Schr\"odinger's equation.
  
\begin{table}[h]
\caption{The ground-state energy of Schr\"odinger's equation (\ref{eq1}) for $\alpha=1$ and different values of $\lambda$ using the present work.  For these computations $r_0=3$. $N$ is the number of iterations.}
{\begin{tabular}{ccc}
\hline
    \multicolumn{3}{c}{$\alpha=1$}  \\
\hline 
 $\lambda$ &  $E_{AIM}$ & N \\
\hline
1000&190.723~307~439~784~825~395~54&90\\
100&42.462~918~114~619~200~840~54&32\\
10&10.577~483~539~371~157~357~99&52\\
1&4.057~877~007~967~971~192~93&64\\
0.1&3.112~066~906~502~466~751~74&65\\
0.01&3.011~276~010~524~898~166~93&62\\
0.001&3.001~128~301~284~079~220~13&60\\
0.0001&3.000~112~837~137~807~781~38&56\\
0.00001&3.000~011~283~783~881~865~84&55\\
0.000001&3.000~001~128~379~089~204~58&54\\
\hline
\end{tabular}}
\end{table}
In Table 4 we report the AIM results for the case $\alpha = 4,$ again for considerable range of $\lambda$. The large number of iteration for small values of $\lambda$ reflect the stability of the method and also the applicability of AIM to treat such cases in one single formalism.  Similar tables can be easily constructed for $\alpha=3,5,6,$ etc. Fewer numbers of iterations are usually needed to achieve any required accuracy for the cases $\gamma>0$. For the ground state energy with $\gamma = 0$, much attention to the value of $r_0$ must be paid to obtain accurate eigenvalues. An important observation is that: for large $n$, the computed roots by means of the termination condition (\ref{eq6}) are either in descending order or in ascending order; if an oscillation is observed, which appears to change the order, then the starting $r_0$ should be revised accordingly. The main point is that with the proper choice of $r_0$, AIM is a stable and efficient method to obtain the eigenenergies to any degree of accuracy.
\section{Conclusion}
The present work points out the importance of the asymptotic wave function used for initializing AIM sequences for Schr\"odinger eigenvalue problems. By introducing a wave function form that satisfies both boundary conditions at zero and at infinity, we were able to obtain accurate eigenvalues for the Schr\"odinger equation with singular potentials. Several suggestions are discussed to remove the numerical instabilities that may be encountered with direct utilization of AIM. Although we have focused our attention on the calculation of eigenenergies, the method also yields the corresponding eigenfunctions via equations (\ref{eq3}) and (\ref{eq4}).   
\section*{Acknowledgments}
\noindent Partial financial support of this work under Grant Nos. GP3438 and GP249507 from the 
Natural Sciences and Engineering Research Council of Canada is gratefully 
acknowledged by two of us (respectively [RLH] and [NS]).
\clearpage  
\begin{table}[h]
\caption{The energy eigenvalues for Schr\"odinger's equation (\ref{eq1}) with $\alpha=4$ and different values of $\gamma$ for a wide range of values of the coupling $\lambda$.  For these computations, $r_0$ varies over the range $[4,6.5]$ as $\lambda$ approaches zero. $N$ is the number of iterations.}
{\begin{tabular}{cccc}
\hline
    \multicolumn{4}{c}{$\alpha=4$}  \\
\hline 
 $\lambda$  &$\gamma$&  $E_{AIM}$ & N \\
\hline
1000&0&21.369~462~532~163~464~497~98&43\\
~&1& 21.522~859~814~112~640~999~87&39\\
~&2& 21.827~883~093~646~909~321~84&40 \\
~&3& 22.281~057~275~014~819~956~75&38  \\
~&4& 22.877~334~674~778~463~023~90&37\\
~&5& 23.610~282~631~878~614~495~76&36\\
\hline
100&0&11.265~080~431~752~838~088~14&68\\
~&1&16.235~741~726~872~875~703~15&69\\
~&2&16.801~763~365~978~787~670~47&67\\
~&3&17.624~891~020~204~476~919~71&65\\
~&4&18.677~251~190~728~170~156~30&63\\
~&5&19.926~538~009~871~777~284~42&58\\
\hline
10&0&~6.606~622~512~024~943~661~69&129\\
~&1&~7.223~520~393~149~576~761~99&124\\
~&2&~8.352~483~528~249~905~501~41&116\\
~&3&~9.839~231~320~856~383~508~94&100\\
~&4&11.544~000~451~519~493~138~54&89\\
~&5&13.371~330~123~959~945~355~25&78\\
\hline
1&0&~4.494~177~983~369~188&275\\
~&1&~5.559~167~225~784~086&246\\
~&2&~7.224~287~163~959~573&158\\
~&3&~9.108~658~607~516~353&131\\
~&4&11.062~241~719~384~166&107\\
~&5&13.040~015~183~057~043&88\\
\hline
0.1&0&~3.575~551~992&260\\
~&1&~5.095~284~821&200\\
~&2&~7.025~961~149&133\\
~&3&~9.011~364~026&90\\
~&4&11.006~336~099&68\\
~&5&13.004~036~433&50\\
\hline
0.01&0&~3.205~067~495&450\\
~&1&~5.011~917~775&356 \\
~&2&~7.002~658~316&155\\
~&3&~9.001~142~200&85\\
~&4&11.000~634~789&60\\
~&5&13.000~404~001&51\\
\hline
0.001&0&~3.068~765~~~~&335\\
~&1&~5.001~286~52&259\\
~&2&~7.000~266~58&88\\
~&3&~9.000~114~28&53\\
~&4&11.000~063~49&39\\
~&5&13.000~040~40&33\\
\hline
\end{tabular}}
\end{table}

\end{document}